\documentclass[aps,prl,twocolumn,superscriptaddress]{revtex4}
\usepackage{amsmath,amssymb,graphicx,subfigure}

\newcommand{\rv}{\mathbf{r}}

\begin{document}

\title{Optical circuits based on Polariton Neurons in Semiconductor Microcavities}

\author{T. C. H. Liew}
\affiliation{School of Physics and Astronomy, University of
    Southampton, Highfield, Southampton SO17 1BJ, UK}
\author{A. V. Kavokin}
\affiliation{School of Physics and Astronomy, University of
    Southampton, Highfield, Southampton SO17 1BJ, UK}
\affiliation{Marie-Curie Chair of Excellence ``Polariton devices",
University of Rome II, 1, via della Ricerca Scientifica, Rome,
00133, Italy}
\author{I. A. Shelykh}
 \affiliation{International Center for Condensed Matter Physics, Universidade de Brasilia, 70904-970, Brasilia-DF, Brazil}
 \affiliation{St. Petersburg State Polytechnical University, 195251, St. Petersburg,  Russia}

\date{\today}

\begin{abstract}
By exploiting the polarization multistability of polaritons, we show
that polarized signals can be conducted in the plane of a
semiconductor microcavity along controlled channels or \lq
neurons\rq. Furthermore due to the interaction of polaritons with
opposite spins it is possible to realize binary logic gates
operating on the polarization degree of freedom. Multiple gates can
be integrated together to form an optical circuit contained in a
single semiconductor microcavity.
\end{abstract}

\maketitle

\noindent Owing partly to their unique spin
structure~\cite{Kavokin2004}, several recent studies have focused on
the spin-dynamics~\cite{Shelykh2005} of exciton-polaritons in
semiconductor microcavities, which operate in the strong
light-matter coupling regime~\cite{Weisbuch1992,CavityPolaritons}.
Exciton-polaritons are part-light, part-matter quasiparticles (that
represent the elementary excitations of the system) and demonstrate
several qualities that make them excellent candidates for practical
devices, including: a long coherence length~\cite{Freixanet2000};
strongly interacting nature~\cite{Savidis2000}; and relative ease of
excitation and detection through coupling to external light fields.
Furthermore, advances in growth technology have resulted in
nanostructures, which give rise to another dimension of
quasiparticle control. In standard planar microcavities, methods
that allow the control of propagating polariton spins have now been
evidenced experimentally~\cite{Leyder2007OSHE} and research has
begun to focus on individual spin optoelectronic
devices~\cite{Shelykh2004,Leyder2007TwoBeam}. A motivation of this
field was to integrate several elements together to create
all-optical circuits, however it was often not clear how such
integration could be achieved.

We demonstrate a technique for making binary logic gates that act on
the polarization degree of freedom of polaritons. Different logic
gates can be linked together, in the plane of the microcavity, as
signals can be carried along controlled channels that are created by
patterning the microcavity structure such that polaritons experience
a structured potential. In fact polaritons themselves do not move
the whole distance from one end of the channel to the other; rather
it is the switching of successive parts of the channel caused by
very short propagation of polaritons that results in a long signal
propagation. In this sense the channel bears a loose analogy to
biological neurons, which is why we call these channels polariton
neurons. Since we do not rely on single polaritons traveling the
full length of the channels, the short lifetime of polaritons does
not limit the length of signal propagation. The logic gates that we
propose rely on the polarization multistability~\cite{Gippius2007}
of polaritons in microcavities.

{\bf Polariton Neurons.} Consider a semiconductor microcavity that
is patterned such that polaritons experience the potential shown in
Fig.~1(b). This can be achieved by variation of the cavity
width~\cite{IdrissiKaitouni2006}, applying stress \cite{Balilli2006}
or putting metals on the surface of the cavity~\cite{Lai2007}. Due
to the long decoherence time~\cite{Langbein2007} and the fact that
in the low density limit they behave as weakly interacting
bosons~\cite{Kasprzak2006}, the dynamics of polaritons can be
treated in the framework of the mean-field approximation, which
leads to the Gross-Pitaevskii (GP)
equation~\cite{Carusotto2004,Shelykh2006}. This equation, commonly
used in the description of atomic Bose-Einstein condensate
dynamics~\cite{GP-BEC}, has been applied to semiconductor
microcavities to describe several phenomena, including: the
suppression of Rayleigh scattering by
impurities~\cite{Carusotto2004}; the spatial structure of
microcavity parametric oscillator polaritons~\cite{Whittaker2005b};
the dispersion of polariton
superfluids~\cite{Shelykh2006,Malpuech2007}; and the interference of
polariton condensates~\cite{Leyder2007TwoBeam}.

Polaritons have two possible spin projections on the structure
growth axis, $\sigma=\pm1$, corresponding to the right ($\sigma_+$)
and left ($\sigma_-$) circular polarizations of external
photons~\cite{Shelykh2005}. The spin dependent GP equation
is~\cite{Gippius2007}:
\begin{align}
&i\hbar\frac{\partial\psi_\sigma}{\partial t}
=\left({\hat{H}_{LP}(-i\hat{\nabla})}-\frac{i\hbar}{2\tau}+W(\rv)\right)\psi_\sigma\notag\\
&+\left(|\psi_\sigma|^2+\frac{\alpha_2}{\alpha_1}|\psi_{-\sigma}|^2\right)\psi_\sigma+p_\sigma(\rv,t)
e^{-i E_p t/ \hbar}, \label{eq:GP}
\end{align}
\noindent where the $\sigma$ polarized internal cavity polariton
field, $\psi_\sigma$, depends on the spatial coordinate, $\rv$. The
kinetic energy operator $\hat{H}_{LP}$ represents the dispersion of
polaritons. We consider only lower branch polaritons from the strong
light-matter coupling - upper branch polaritons will not be excited
under the conditions we propose. $\tau$ is the polariton lifetime.
$W(\rv)$ represents a potential experienced by polaritons.
$\alpha_{1(2)}$ is the matrix element of polariton-polariton
interaction in the parallel spin(antiparallel spin) configuration,
respectively. It is well known that for 2D excitons and
exciton-polaritons the exchange interaction strongly dominates over
the direct one, and thus polariton-polariton interactions are
anisotropic $|\alpha_2|< \alpha_1$~\cite{Exchange}. This anisotropy
strongly affects the properties of polariton systems in the
superfluid regime~\cite{Shelykh2006,Rubo2006} and leads to
remarkable nonlinear effects in polariton spin relaxation, such as
self-induced Larmor precession and inversion of the linear
polarization during the scattering act~\cite{Shelykh2005,
Krizhanovskii2006}. In Eq.~\ref{eq:GP} the fields were rescaled so
that only the ratio of $\alpha_2$ to $\alpha_1$ is significant.

    \begin{figure}[h]
        \centering
            \subfigure{\includegraphics[width=4.058cm]{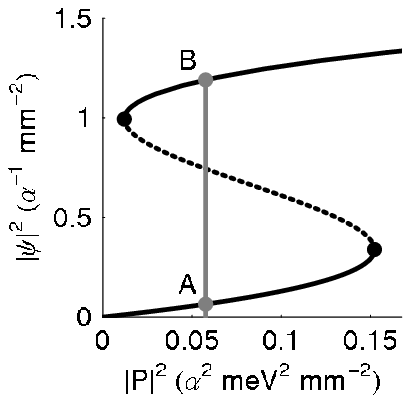}\put(-170,110){\parbox{4.058cm}{\sf \bf a}}}
            \subfigure{\includegraphics[width=4.058cm]{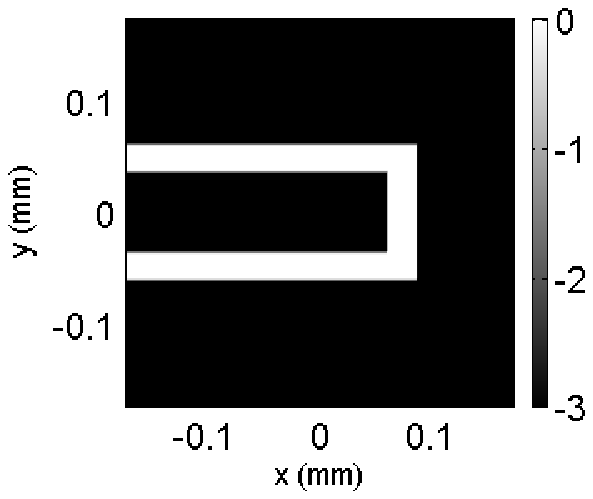}\put(-170,110){\parbox{4.058cm}{\sf \bf b}}}
    \caption{a) Dependence of the intensity of a single uncoupled (spin-polarized) polariton state on pump power.
    The S-shaped curve is characteristic of a bistable system~\cite{Baas2004,Gippius2004,Whittaker2005} and it is
    well-known that the middle branch (dotted part of curve) is unstable, that is, the polariton intensity can have a maximum of two
    possible values, A \& B, for a given pump intensity. Parameters: $E_p-E_0=1$meV, $\tau=3$ps. b) Polariton potential profile in real space.}
    \label{fig:SCurve}
    \end{figure}

The driving optical pump field is given by $p_\sigma(\rv,t)$ and
$E_p$ is the pump energy. If the pump energy is tuned greater than
$\hbar\sqrt{3}/\tau$ above the polariton eigenenergy (bare polariton
branch energy) then for some excitation powers the system can
exhibit more than one stable
state~\cite{Whittaker2005,Baas2004,Gippius2004}. The dependence of
the polariton intensity of a single state in space can be calculated
analytically~\cite{Gippius2007} from the GP equation in the
stationary regime, if coupling to other points in space is ignored
(i.e., one assumes an infinite polariton effective mass):
\begin{equation}
\left[\left(E_0-E_p+|\psi_\sigma|^2+\frac{\alpha_2}{\alpha_1}|\psi_{-\sigma}|^2\right)^2+\frac{\hbar^2}{4\tau^2}
\right]|\psi_\sigma|^2=|p_\sigma|^2, \label{eq:Bistability}
\end{equation}
where $E_0$ is the bare polariton eigenenergy. If, for simplicity we
consider the excitation of the system by circularly polarized light
then all polaritons will have the same spin and the polariton
intensity exhibits an S-shaped curve, which characterizes a bistable
system [Fig.~1(a)]. If the pump intensity is increased from zero,
the polariton intensity increases steadily from zero until the first
turning point. For higher pump intensities, the polariton intensity
jumps to the upper branch of the S-shaped curve. If the pump
intensity is then decreased, the polariton intensity remains on this
branch, provided the pump intensity is greater than that of the
second turning point.

We solve Eq.~\ref{eq:GP} numerically (with finite effective
polariton mass) to model the excitation of the system in Fig.~1(b)
with a broad, $\sigma_+$ polarized, Gaussian, cw pump. The pump is
tuned above the polariton eigenenergy of the channel region and has
weak intensity such that polariton intensities lie on the lower
branch of the S-shaped curve. Due to the larger pump-eigenenergy
detuning, hardly any polaritons are excited in the region outside
the channel. We then calculate the evolution of the polariton fields
after a ($\sigma+$ polarized) pulse is applied near one end of the
channel, which locally switches the polariton intensity to the upper
branch of the S-shaped curve.

    \begin{figure}[h]
        \centering
                \includegraphics[width=4.058cm]{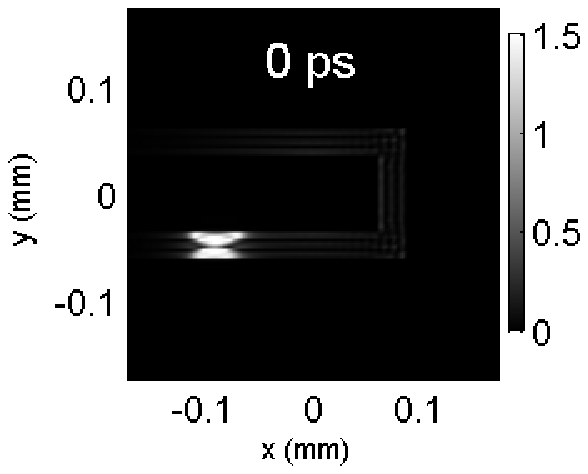}
                \includegraphics[width=4.058cm]{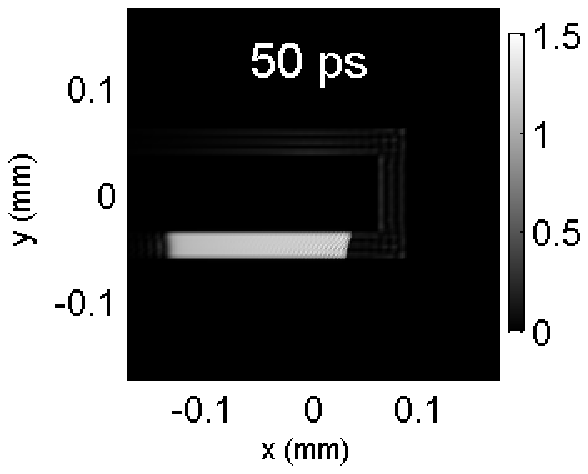}
                \\
                \includegraphics[width=4.058cm]{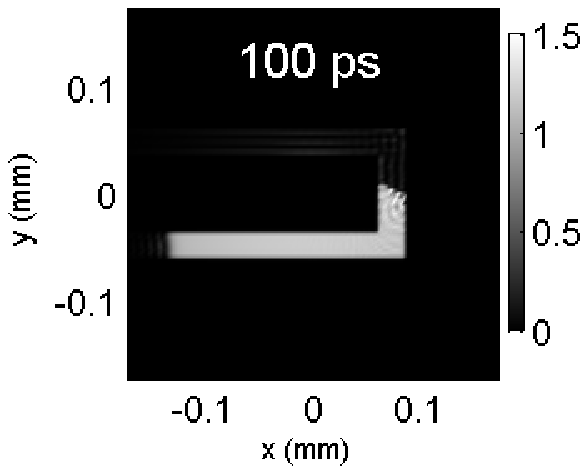}
                \includegraphics[width=4.058cm]{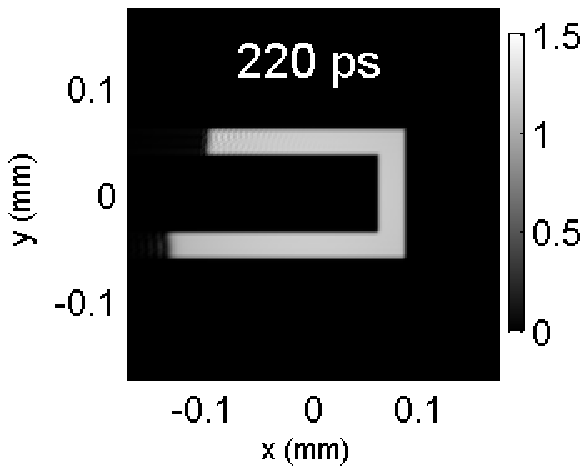}
    \caption{Spatial polariton intensity profile at different times relative to the pulse arrival
    time. The polariton lifetime was $\tau=3ps$ and the polariton dispersion was calculated with a two oscillator model in which the exciton-photon coupling energy was $3$meV and the photon effective mass was $\times10^{-5}$ the free electron mass. Both the cw pump and pulse correspond to optical fields at normal incidence, that is, they have Gaussian distributions in reciprocal space centered at zero in-plane wavevector. The pumps are tuned $1$meV above the bare polariton eigenenergy.}
    \label{fig:Neuron}
    \end{figure}

The results (Fig.~2) show that after the pulse has decayed,
successive regions of the channel switch to the high intensity
state. This propagation of the signal continues around corners in
the channel, allowing a way of creating wires for optical circuits
in the microcavity plane.
% We note that in principle the containment of polaritons in channels could also be achieved by patterning the weak driving cw field instead of structuring the potential.
The switching mechanism is analogous to the optical switching waves
observed in bistable semiconductor microresonators with fast
electronic nonlinearity~\cite{Ganne2005}.

The signal propagation speed depends strongly on the intensity of
the driving cw field. For our parameters it is $1.8\times10^6$ m/s.
The propagation speed could be enhanced by driving the system with a
finite in-plane wavevector - however this would introduce a
dependence of the signal propagation speed on the neuron in-plane
direction.

{\bf Logic Gates.} Logic gates, for creating optical circuits in the
microcavity plane, can be created by exploiting the polarization
degree of freedom of polaritons. When considering the polarization
degree of freedom, our system demonstrates a
multistability~\cite{Gippius2007} rather than bistability; both the
$\sigma_+$ and $\sigma_-$ polariton field intensities can exhibit
the S-shaped curve of Fig.~1, and both can lie either on the upper
or lower branch of the curve.

We now consider the merging of two polariton neurons, by
restructuring the potential profile. The system is again excited by
a broad but weak Gaussian cw pump, which is now elliptically
polarized with a bias towards the $\sigma_+$ polarization. The two
input channels are independently excited by either $\sigma_+$ or
$\sigma_-$ pulses. In Fig.~3 we plot the circular polarization
degree in real space for the case of oppositely polarized inputs
(left column) and the case of two $\sigma_-$ inputs (right column).
The circular polarization degree is defined as
$\rho_c=\frac{|\psi_+|^2-|\psi_-|^2}{|\psi_+|^2+|\psi_-|^2}$.

    \begin{figure}[h]
        \centering
        \includegraphics[width=4.058cm]{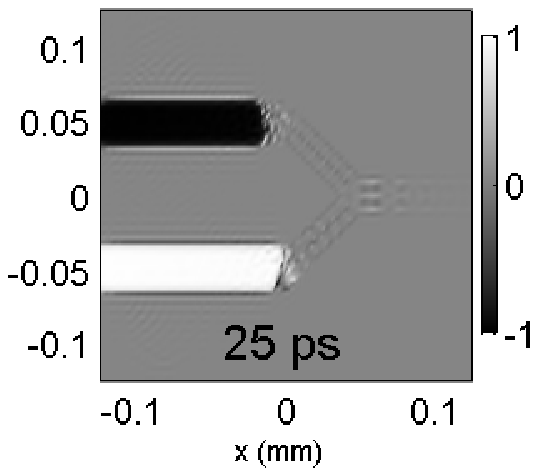}
        \includegraphics[width=4.058cm]{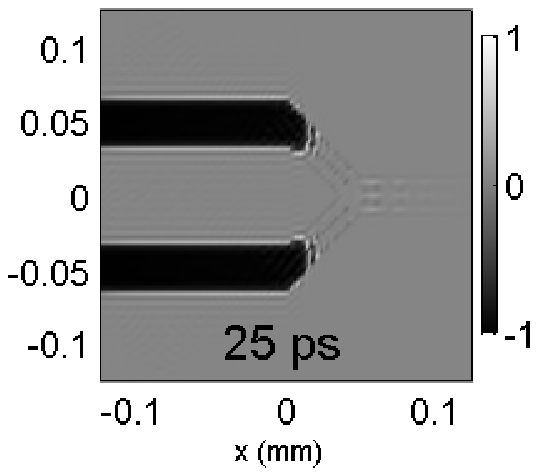}
        \\
        \includegraphics[width=4.058cm]{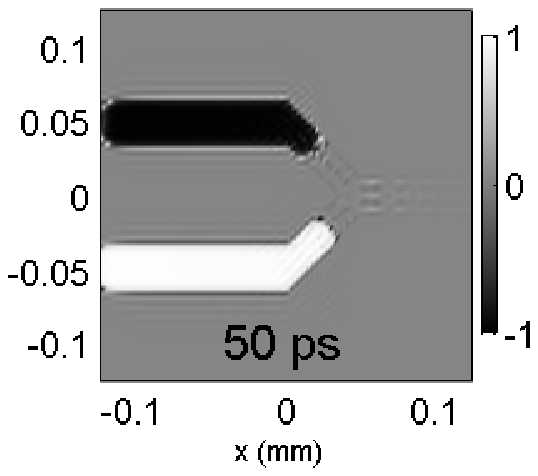}
        \includegraphics[width=4.058cm]{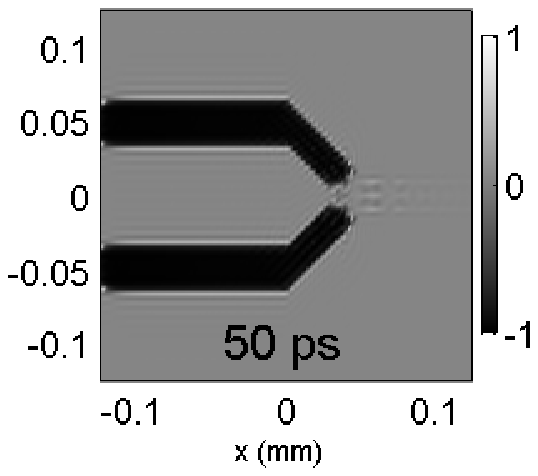}
        \\
        \includegraphics[width=4.058cm]{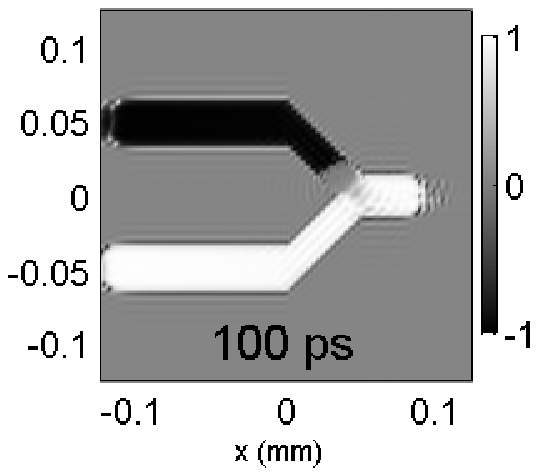}
        \includegraphics[width=4.058cm]{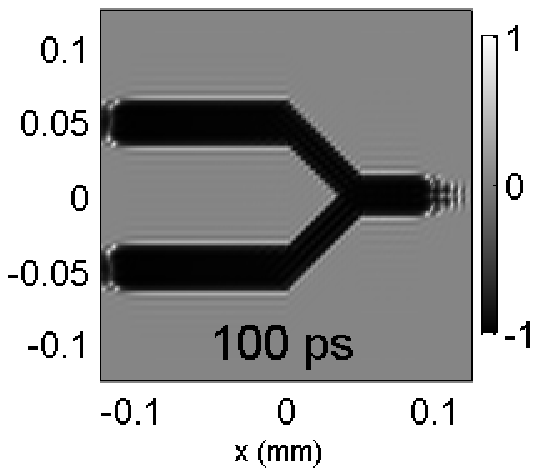}

    \caption{Circular polarization degree, $\rho_c$, in a region where two polariton neurons merge, at different times relative to the pulse
    arrival time. In the left column the system is excited with
    oppositely circularly polarized pulses; in the right column it is excited with two $\sigma_-$ polarized pulses. $\alpha_2=-0.5\alpha_1$.}
    \label{fig:Gate}
    \end{figure}

When the signals of polariton neurons firing with opposite spin
polarization overlap, we find that only the $\sigma_+$ signal
continues. This is due to the bias of the cw background field toward
$\sigma_+$ and a negative value of $\alpha_2$. This means that the
system behaves as an OR logic gate from which the output is
$\sigma_+$ polarized if either input is $\sigma_+$ polarized
(alternatively we would have an AND gate if the background cw field
had opposite $\rho_c$).

For the parameters used in our device an uncertainty in the arrival
time of pulses of at least $10$ps is allowed (note that this is
$10$\% of the device operation time). The uncertainty can be
estimated as the difference in arrival time at the junction of
$\sigma_+$ and $\sigma_-$ polarized signals when triggered
simultaneously. The allowed uncertainty can be increased by
adjusting the relative intensities of the cw $\sigma_+$ and
$\sigma_-$ components (to alter the propagation speed of $\sigma_+$
and $\sigma_-$ signals) or by increasing the distance between the
excitation points and the junction.

The power requirements of the cw optical pump would make up the
majority of the power consumption of the device. We note that in
Ref.~\onlinecite{Baas2004}, histeresis was observed in GaAs based
microcavities using a power of $2.8$mW. In microcavities with a
larger exciton-exciton interaction strength (e.g., GaN based
microcavities suitable for room temperature operation) lower power
requirements could be achieved. Furthermore the required pump power
is sensitive to the polariton lifetime, which can be increased by
using higher Q-factor cavities.

{\bf Conclusion.} In conclusion, the short range propagation of
polaritons enables the successive switching of neighboring regions
along a multistable channel in space from a low intensity stable
state to a high intensity stable state. This switching allows a
signal to continue over distances longer than the distance a single
polariton could travel before it decays (the distance is limited by
the extent of the background cw field, which must have sufficient
intensity at a given point for a multistability to exist). When
channels merge the polarization of the output propagating signal
depends on the ingoing polarizations in a logical way, which allows
the construction of binary logic gates. This gives us the
opportunity to build optical circuits, in which multiple elements
are integrated within a single microcavity structure.

I.A.S. acknowledges the support from the Grant of the President of
Russian Federation. T.C.H.L acknowledges support from the E.P.S.R.C.


\begin{thebibliography}{10}

\bibitem{Kavokin2004}
%K. V. Kavokin, I. A. Shelykh, A. V. Kavokin, G. Malpuech, P. Bigenwald
K. V. Kavokin, {\it et. al.}, Phys. Rev. Lett., {\bf 92}, 017401
(2004).

\bibitem{Shelykh2005}
I. A. Shelykh, A. V. Kavokin, G. Malpuech, Phys. Stat. Sol. (b),
{\bf 242}, 2271 (2005).

\bibitem{Weisbuch1992}
C. Weisbuch, M. Nishioka, A Ishikawa, Y. Arakawa, Phys. Rev. Lett.,
{\bf 69}, 3314 (1992).

\bibitem{CavityPolaritons}
A. V. Kavokin, G. Malpuech,
\newblock {\it Cavity Polaritons},
\newblock Elsevier (2003).

\bibitem{Freixanet2000}
T. Freixanet, B. Sermage, A. Tiberj, R. Planel, Phys. Rev. B, {\bf
61}, 7233 (2000).

\bibitem{Savidis2000}
% P. G. Savvidis, J. J. Baumberg, R. M. Stevenson, M. S. Skolnick, D. M. Whittaker, J. S. Roberts
P. G. Savvidis, {\it et. al.}, Phys. Rev. Lett., {\bf 84}, 1547
(2000);

\bibitem{Leyder2007OSHE}
%C. Leyder, T. C. H. Liew, A. V. Kavokin, I. A. Shelykh, M. Romanelli, J. Ph. Karr, E. Giacobino, A. Bramati
C. Leyder, {\it et. al.}, Nat. Phys., {\bf 3}, 628 (2007).

% Semiconductor microcavity as a spin-dependent optoelectronic device
\bibitem{Shelykh2004}
%I. Shelykh, K. V. Kavokin, A. V. Kavokin, G. Malpuech, P. Bigenwald, H. Deng, G. Weihs, Y. Yamamoto,
I. Shelykh, {\it et. al.}, Phys. Rev. B, {\bf 70}, 035320 (2004).

\bibitem{Leyder2007TwoBeam}
%C. Leyder, T. C. H. Liew, A. V. Kavokin, I. A. Shelykh, M. Romanelli, J. Ph. Karr, E. Giacobino, A. Bramati,
C. Leyder, {\it et. al.}, Phys. Rev. Lett., {\bf 99}, 196402 (2007).

\bibitem{Gippius2007}
%N. A. Gippius, I. A. Shelykh, D. D. Solnyshkov, S. S. Gavrilov, Yuri G. Rubo, A. V. Kavokin, S. G. Tikhodeev, G. Malpuech,
N. A. Gippius, {\it et. al.}, Phys. Rev. Lett., {\bf 98}, 236401
(2007).

\bibitem{IdrissiKaitouni2006}
%R. Idrissi Kaitouni, O. El Da\"{i}f, A. Baas, M. Richard, T. Paraiso, P. Lugan, T. Guillet, F. Morier-Genoud, J. D. Gani\`{e}re, J. L. Staehli, V. Savona, B. Deveaud,
R. Idrissi Kaitouni, {\it et. al.} Phys. Rev. B, {\bf 74}, 155311
(2006).

\bibitem{Balilli2006}
R. B. Balili, D. W. Snoke, L. Pfeiffer, K. West, Appl. Phys. Lett.,
{\bf 88}, 031110 (2006).

\bibitem{Lai2007}
%C. W. Lai, N. Y. Kim, S. Utsunomiya, G. Roumpos, H. Deng, M. D. Fraser, T. Byrnes, P. Recher, N. Kumada, T. Fujisawa, Y. Yamamoto,
C. W. Lai, {\it et. al.}, Nature, {\bf 450}, 529 (2007).

\bibitem{Langbein2007}
%W. Langbein, I. Shelykh, D. Solnyshkov, G. Malpuech, Yu. Rubo, A. Kavokin,
W. Langbein, {\it et. al.}, Phys. Rev. B, {\bf 75}, 075323 (2007).

\bibitem{Kasprzak2006}
The Bose condensation of polaritons signifies their bosonic
character and has been reported by:
%J. Kasprzak, M. Richard, S. Kundermann, A. Baas, P. Jeambrun, J. M. J. Keeling, F. M. Marchetti, M. H. Szyma\'{n}ska, R. Andr\'{e}, J. L. Staehli, V. Savona, P. B. Littlewood, B. Deveaud, Le Si Dang,
J. Kasprzak, {\it et. al.}, Nature {\bf 443}, 409 (2006);
%R. Balili, V. Hartwell, D. Snoke, L. Pfeiffer, K. West,
R. Balili, {\it et. al.}, Science, {\bf 316}, 1007 (2007); see also
Ref.\cite{Lai2007}.

% Polarization & Propagation of Polariton Condensates
\bibitem{Shelykh2006}
%I. A. Shelykh, Yuri G. Rubo, G. Malpuech, D. D. Solnyshkov, A. Kavokin,
I. A. Shelykh, {\it et. al.}, Phys. Rev. Lett., {\bf 97}, 066402
(2006).

\bibitem{Carusotto2004}
I. Carusotto, C. Ciuti, Phys. Rev. Lett., {\bf 93}, 166401(2004).

\bibitem {GP-BEC}
F. Dalfovo, Stefano Giorgini, Lev P. Pitaevskii, Sandro Stringari,
Rev. Mod. Phys., {\bf 71}, 463 (1999); A.J. Legett, Rev. Mod. Phys.,
{\bf 73}, 307 (2001);
%L. Pitaevskii, S. Stringari, Bose-Einstein
%Condensation, ISBN-13: 978-0198507192, Oxford University press,
%2003.

\bibitem{Whittaker2005b}
D M Whittaker, Phys. Stat. Sol. (c), {\bf 2}, 733 (2005);
%D.Sanvitto, D. N. Krizhanovskii, D. M. Whittaker, S. Ceccarelli, M. S. Skolnick, J. S. Roberts,
D.Sanvitto, {\it et. al.}, Phys. Rev. B {\bf 73}, 241308 (2006).

\bibitem{Malpuech2007}
%G. Malpuech, D. D. Solnyshkov, H. Ouerdane, M. M. Glazov, I. Shelykh,
G. Malpuech, {\it et. al.}, Phys. Rev. Lett. {\bf 98}, 206402
(2007).

\bibitem{Exchange}
%C. Ciuti, V. Savona, C. Piermarocchi, A. Quattropani, P. Schwendimann,
C. Ciuti, {\it et. al.}, Phys. Rev. B, {\bf 58}, 7926 (1998); M.
Combescot \& O. Betbeder-Matibet,  Phys. Rev. B, {\bf 74}, 125316
(2006);
%P. Renucci, T. Amand, X. Marie, P. Senellart, J. Bloch, B. Sermage, K. V. Kavokin, Phys. Rev. B, {\bf 72}, 075317 (2005).
P. Renucci, Phys. Rev. B, {\bf 72}, 075317 (2005).

\bibitem{Rubo2006}
%Yu. G. Rubo, A. V. Kavokin, I. A. Shelykh,
Yu. G. Rubo, {\it et. al.}, Phys. Lett. A, {\bf 358}, 227 (2006);
I.A. Shelykh, Yu.G. Rubo, A.V. Kavokin, Superlattices Microstruct.,
{\bf 41}, 313 (2007).

\bibitem{Krizhanovskii2006}
%D.N. Krizhanovskii, D. Sanvitto, I. A. Shelykh, M. M. Glazov, G. Malpuech, D. D. Solnyshkov, A. Kavokin, S. Ceccarelli, M. S. Skolnick, J. S. Roberts,
D. N. Krizhanovskii, {\it et. al.}, Phys. Rev. B, {\bf 73}, 073303
(2006).

\bibitem{Whittaker2005}
D. M. Whittaker, Phys. Rev. B, {\bf 71}, 115301 (2005).

\bibitem{Baas2004}
A. Baas, J. Ph. Karr, H. Eleuch, E. Giacobino, Phys. Rev. A, {\bf
69}, 023809 (2004).

\bibitem{Gippius2004}
%N. A. Gippius, S. G. Tikhodeev, V. D. Kulakovskii, D. N. Krizhanovskii, A. I. Tartakovskii,
N. A. Gippius, {\it et. al.}, Europhys. Lett., {\bf 67}, 997 (2004).

\bibitem{Ganne2005}
I. Ganne, G. Slekys, I. Sagnes, R. Kuszelewicz, Phys. Rev. B, {\bf
63}, 075318 (2001).

\end{thebibliography}
\end{document}